\documentclass[a4paper,fleqn,usenatbib,letters]{mnras}

\usepackage{graphicx}	
\usepackage{amsmath}	
\usepackage{amssymb}	
\usepackage{physics}	
\usepackage{bm}			

\newcommand{\sepparam}{s_0}

\renewcommand{\phi}{\varphi}

\newcommand{\imparam}{u_0}

\newcommand{\vsource}{\mu_{rel}}

\newcommand{\inc}{\iota}

\newcommand{\truanom}{f}

\renewcommand{\vec}{\bm}

\newcommand{\change}{\textcolor{black}}

\title[Detecting binarity in BBH microlensing events]{Detecting binarity of GW150914-like lenses in gravitational microlensing events}

\author[D.H. Eilbott et al.]{
Daniel H. Eilbott\thanks{Email contact: daniel.eilbott@utdallas.edu (DHE); ~ ~ ~ ~ ~ ~ ~ ~ ~ kesden@utdallas.edu (MK); lindsay.king@utdallas.edu (LJK)}, 
Alexander H. Riley, Jonathan H. Cohn, Michael Kesden\footnotemark[1], \newauthor ~and Lindsay J. King\footnotemark[1]
\\
Department of Physics, The University of Texas at Dallas, 800 W Campbell Rd, Richardson, TX 75080, USA}

\date{\today}

\begin{document}
\label{firstpage}
\pagerange{\pageref{firstpage}--\pageref{lastpage}}
\maketitle

\begin{abstract}
The recent discovery of gravitational waves (GWs) from stellar-mass binary black holes (BBHs) provided direct evidence of the existence of these systems. BBH lenses would have gravitational microlensing signatures that are distinct from single-lens signals. We apply Bayesian statistics to examine the distinguishability of BBH microlensing events from single-lens events under ideal observing conditions, using the photometric capabilities of the Korean Microlensing Telescope Network. Given one year of observations, a source star at the Galactic Centre, a GW150914-like BBH lens (total mass $65M_\odot$, mass ratio 0.8) at half that distance, and an impact parameter of 0.4 Einstein radii, we find that binarity is detectable for BBHs with separations down to 0.0250 Einstein radii, which is nearly 3.5 times greater than the maximum separation for which such BBHs would merge within the age of the Universe. Microlensing searches are thus sensitive to more widely separated BBHs than GW searches, perhaps allowing the discovery of BBH populations produced in different channels of binary formation.
\end{abstract}


\section{\label{intro}Introduction}
The 2015 discovery (\cite{ligo}) of gravitational waves (GWs) provided the first direct proof of the existence of stellar-mass binary black holes (BBHs). This paper investigates the feasibility of identifying such systems in our Galaxy through gravitational microlensing. BBH systems with the potential to produce detectable GWs have sufficiently small separations to complete hundreds of orbits over the course of a lensing event.

This realm of microlensing, where orbital motion of the lens is detectable, is well explored. \change{\cite{zheng} discussed the effects of superluminal caustics due to the rapid rotation of tight binary lenses. \cite{dubath} examined the contribution of the quadrupole of compact binary lenses to microlensing signals. \cite{pennya} examined the detection efficiency of orbital motion in binary microlensing events with varying parameters, while \cite{pennyb} extended this investigation to include binary lenses that show at least one repeating feature in their light curve. \cite{nucita} studied the use of orbital motion signals in microlensing to identify the binary lens period. \cite{sajadian2014} and \cite{sajadian2015} explored the efficiency of detecting binary astrometric microlensing signals.} Our goal is to apply Bayesian statistics to the binary microlens to examine whether BBH microlensing signals from within our Galaxy could be discernible with current telescopes under ideal observing conditions (i.e. high cadence over the course of the event and no extinction).

This work is structured as follows: In Section 2, we develop our BBH lensing model. In Section 3, we apply Bayesian statistics to constrain detectability. We present our results in Section 4 and our conclusions in Section 5.

\section{\label{methods}Methods}
\begin{figure*}
\includegraphics[width=\textwidth]{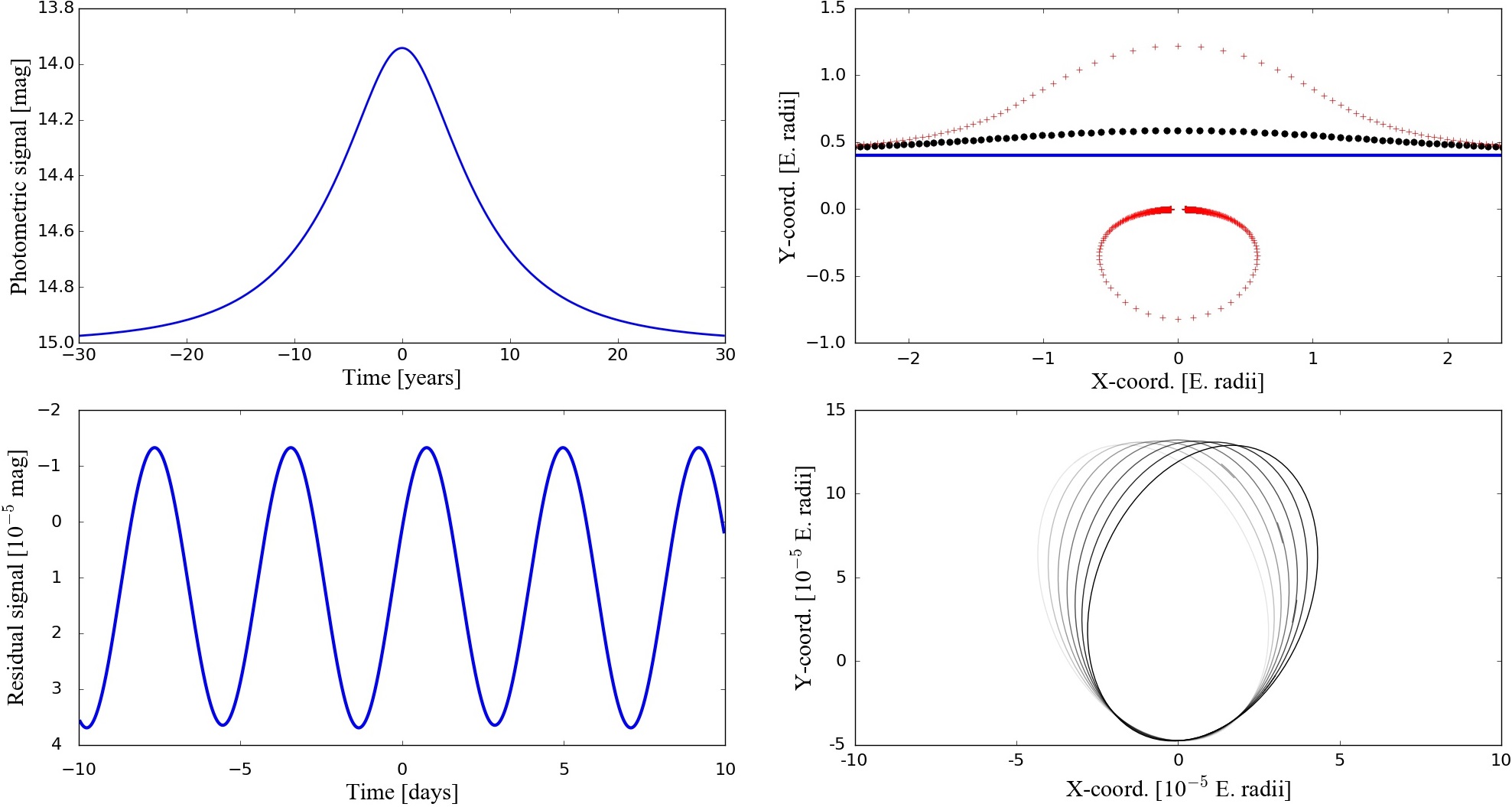}
\caption{Top left: The BBH lens photometric light curve magnification. Bottom left: The residual light curve, centred at the event peak, calculated by subtracting the single lens signal from the binary lens signal. The scale of the residual is $10^{-5}$ that of the light curve above it. Top right: The BBH lens source path (blue line), image paths (red +), and centroid position (black circles), with the BBH centre of mass at the origin. While three images are produced in the event, the third is highly de-magnified and therefore negligible for our analysis. Bottom right: seven sample curves of the residual centroid, each corresponding to the residual centroid shift during one period of the BBH orbit. The curves are sampled approximately 50 days apart, with the fourth curve corresponding to the peak of the event. The shading moves from pale to dark as time progresses during the event. All four plots use our fiducial model: total mass $M = 65M_{\odot}$, mass ratio $q = 0.8$, binary separation $\sepparam = 0.01$ Einstein radii, \textcolor{black}{proper motion $\vsource=5.5$ mas yr$^{-1}$}, impact parameter $\imparam = 0.4$ Einstein radii, inclination $\inc = \arccos{0.6}$, true anomaly $\truanom = 1$ rad, and direction of proper motion $\phi = 0.3$ rad.} 
\end{figure*}
The binary lens has been explored in detail (e.g. see \cite{schneider}, \cite{stefano}, and \cite{han}). The lensing equation generalised for $N$ point lenses is (e.g. \cite{dominik}):
\begin{equation}\label{nlens}
\vec{y} = \vec{x} - \sum_{r=1}^{N}m_r\dfrac{\vec{x} - \vec{x}^{(r)}}{|\vec{x} - \vec{x}^{(r)}|^2}
\end{equation}
where $\vec{y}$ refers to the source position in the source plane normalized to the Einstein radius, $\vec{x}$ corresponds to the normalized image position in the lens plane, $\vec{x}^{(r)}$ corresponds to the normalized position of the $r$th lens, and $m_r$ are the mass fractions of the lensing objects. We use \eqref{nlens} for the binary case, where the two black holes are assumed to travel on circular orbits about their centre of mass.

Our BBH system is described by eight parameters: the total mass $M$ of the black holes, the mass ratio $q$ of the black holes, the binary separation $\sepparam$ (normalized to the Einstein radius), the impact parameter $\imparam$ (normalized to the Einstein radius), the proper motion $\vsource$ of the source relative to the lens, the inclination $\inc$, the true anomaly $\truanom$ at the closest approach of the source, and the angle $\phi$ between the proper motion direction and the major axis of the BBH orbit's projected ellipse.

The BBHs of interest here have separations much smaller than their Einstein radius and generally much smaller than the impact parameter. Given such small separations, caustic crossings---whether superluminal, such as those discussed in \cite{zheng}, or otherwise---are unlikely (see \cite{dominik}) and we therefore assume that the source star is a point source, as well as that three images are produced rather than five (though we exclude one of the images in our analysis due to its extreme faintness). BBHs of small binary separation will also have orbital periods much shorter than the duration of the lensing event. Thus, their microlensing light curves will consist of a quasi-periodic perturbation superimposed on the standard curve produced by a single lens of the same total mass (e.g. \cite{guo}). Our fiducial model corresponds to an orbital period of 8.4 days; therefore, the quasi-symmetric period of the lensing signal will be 4.2 days.

Using the eight model parameters, we find the projected lens and source positions.
The partial derivatives of our modified equations comprise a Jacobian matrix; inverting the determinant of the Jacobian matrix yields the image magnifications. We iterate through the lensing event, calculating the image positions and magnifications numerically. Our model assumes exposure times that are shorter than the binary period. 

The analytical image positions of a single lens with corresponding system parameter values provide initial guesses for our numerical approximation. The binary lens's quasi-sinusoidal deviation from the standard single lens light curve is small compared to the curve itself, so the single lens image positions are an efficient means of initiating the solver. As expected (see \cite{dominik}), the amplitudes of the binary lens perturbations are proportional to the quadrupole moment $\eta M {\sepparam}^2$, where $\eta \equiv q/(1+q)^2$. Fig.~1 shows plots of binary and residual light curves, the binary image paths and centroid, and the residual centroid. Note that the \change{Einstein} ring crossing timescale of the lensing event for our fiducial model parameters (see Fig.~1 caption) is $t_E \equiv \theta_E/\vsource = 1.48$ yr.

\section{Model Selection}
\subsection{\label{bayes}Bayesian statistics}
For a binary lens to be distinguishable from a single lens, the size of the perturbations due to the binary lens must be detected with statistical certainty. Towards this end, we employ Bayesian statistics (e.g. see \cite{stats}). \cite{trotta} described the advantages and disadvantages of employing Bayesian rather than frequentist statistics.

The Bayesian likelihood, $L \equiv \exp\left(-\chi^2 / 2\right)$, measures how closely a model fits data:
\begin{equation}\label{like}
\chi^2 = \sum_{k} \dfrac{(d_k - m_k)^2}{{\sigma_k}^2}\,.
\end{equation}

Here we sum over $k$ points, where $d_k$ is the observed event signal measured in number of photons, $m_k$ is the modelled result (also in photons),  and $\sigma_k$ represents the error associated with each measured point. In lieu of real observations, the output of the lensing code with event parameters serves as the signal. As such, for a model with parameters equal to those used to generate the event signal, all $m_k = d_k$ and $L = 1$. 

The Bayesian evidence $E$ is used for model selection. The evidence is the integral  of the likelihood multiplied by the prior $p$ over all parameter space (e.g. \cite{stats}):
\begin{equation} \label{evidence}
E = \int Lp\,d\vec{\xi}\,.
\end{equation}

With our 8-parameter binary model, numerically evaluating the integral in \eqref{evidence} is computationally expensive. Therefore, we use the Fisher matrix to approximate the likelihood (e.g. \cite{hobson}).

\subsection{\label{fisher}Fisher matrix approximation}
Using exposure times shorter than the binary period, the appropriate Fisher matrix is:
\begin{gather}
F_{ij} = \sum_k\dfrac{1}{{\sigma_k}^2} \pdv{m_k}{\xi_i} \pdv{m_k}{\xi_j} \label{photofisher}
\end{gather}
where $\vec{\xi}$ is the set of parameters, $\sigma_k$ are the photometric errors associated with the $k$th point, and $\pdv{m}{\xi}$ are numerical derivatives of the photometric lensing curve.

The resulting real symmetric 8$\times$8 Fisher matrix is used to approximate the likelihood. Taking $m_k(\vec{\xi}) \simeq m_k (\vec{\xi}_0) + \sum_j\pdv{m_k}{\xi_j} (\xi_j - \xi_{0j})$, we find:
\begin{align}
\chi^2(\vec{\xi}) &= \sum_i\sum_j\sum_k \frac{1}{{\sigma_k}^2} \pdv{m_k}{\xi_i} \pdv{m_k}{\xi_j} (\xi_i - \xi_{0i}) (\xi_j - \xi_{0j}) \\
&= \sum_i\sum_j F_{ij}(\xi_i - \xi_{0i})(\xi_j - \xi_{0j})
\end{align}
where $\xi_0$ is the fiducial parameter set. Comparing the numerical likelihood with the Fisher matrix approximation, we confirm that the approximation is valid near the peak, which will dominate the evidence integral in \eqref{evidence}.

We consider hypothetical observations with the photometric accuracy and cadence of the Korean Microlensing Telescope Network (KMTNet). KMTNet's 10 minute cadence provides many samples of the light curve over the course of one binary period. From equations (43) through (47) of \cite{henderson14}:
\begin{equation} \label{photoerr}
\sigma_k = \sqrt{N_\text{obj} + \left(N_\text{obj}\sigma_\text{sys}\dfrac{\ln{10}}{2.5}\right)^2}
\end{equation}
where $\sigma_\text{sys} = 0.004$ magnitudes is the systematic fractional error floor and $N_\text{obj} = 4.91\times t_\text{exp}\times 10^{-0.4(\text{I} - 22.0)}$ is the number of photons measured from the lensed source star. $t_\text{exp}$ is the 120 second exposure time and $I$ is the lensed star's apparent magnitude in the $I$ band. We assume the sky background is negligible.

We do not consider astrometric information in our analysis, as there are no current astrometric surveys with high enough cadence to effectively detect the small changes in centroid position. For example, \text{Gaia} only expects $\sim14$ measurements per year (see \cite{esa}).

\subsection{\label{priors}Priors and evidences}
To calculate the Bayesian evidence, we set a prior on each of our parameters. The prior for $M$ is flat from $2M_{\odot}$ to $200M_{\odot}$. Note that the black hole masses identified in the first two detected BBH mergers are $36M_{\odot}$ and $29M_{\odot}$ (\cite{ligo}) and $14M_{\odot}$ and $8M_{\odot}$ (\cite{ligo2}). The prior for $q$ is flat from 0.1 to 1. Our BBH separation prior is flat in $\log{\sepparam}$ from $-4$ to $-1$, with $\sepparam$ in units of Einstein radii. Uncertainties in the stellar evolution of BBH progenitors yield poor constraints on BBH masses and separations making the choice of flat priors a reasonable one (see \cite{ligo3}). For $\imparam$, the prior is flat from 0.3 to 2 Einstein radii. Since the size of a tight binary lens's central caustic is proportional to ${\sepparam}^2$ (see \cite{dominik}), the chances of a caustic crossing, given our small BBH separations and allowed values of $\imparam$, are low. For $\vsource$, our prior is flat between 0 and 10 mas yr$^{-1}$ (see \cite{henderson15}). The priors for $\phi$ and $\truanom$, respectively, are flat from 0 to $\pi$ and 0 to $2\pi$ radians. For inclination, the prior is flat in $\cos{\inc}$ from -1 to 1. \change{Note that, while flat priors for some of our parameters are unrealistic, the assumptions are reasonable given poor constraints and the resulting ease of computation.}

Since all of the priors are flat, the prior term can be pulled out of the integral in \eqref{evidence} as a factor of $p = V^{-1}$, where $V \equiv \prod_i \left({\xi_i}^+ - {\xi_i}^-\right)$ is the volume of the priors. ${\xi_i}^+$ is the upper boundary of the prior for parameter $\xi_i$ and ${\xi_i}^-$ is the lower boundary. Using the Fisher matrix approximation, the evidence is then:
\begin{equation} \label{evid}
E = \dfrac{1}{V} \int\exp\left(-\dfrac{1}{2}\sum_i\sum_j F_{ij}(\vec\xi - \vec\xi_0)_i(\vec\xi - \vec\xi_0)_j\right)d\vec\xi\,.
\end{equation}
We use Monte Carlo integration to compute the evidence integrals.
\begin{figure*}
\includegraphics[width=\textwidth]{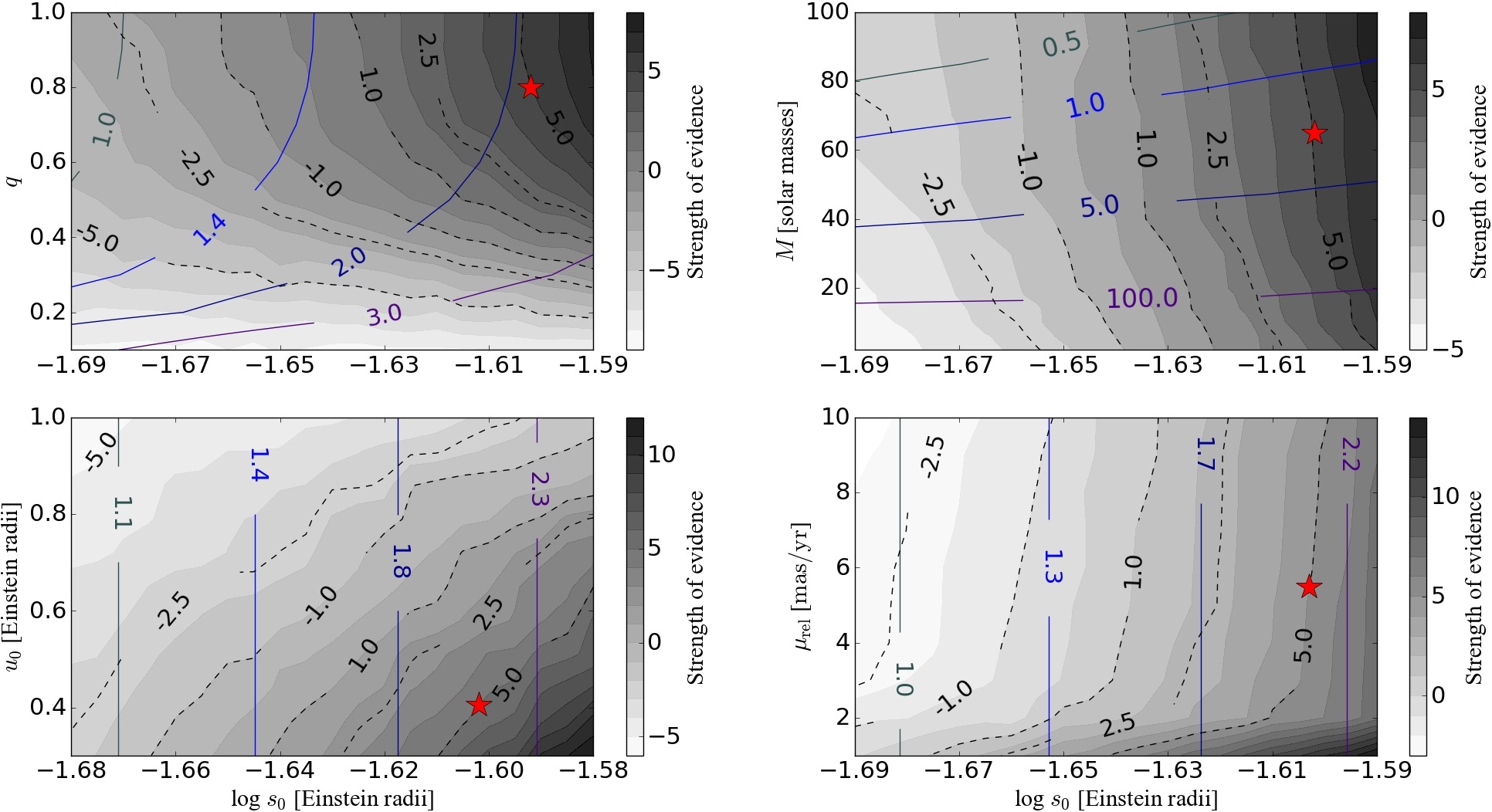}
\caption{Contours of the log Bayes factor $\ln{B}$ for our fiducial model: total mass $M = 65M_{\odot}$, mass ratio $q = 0.8$, proper motion $\vsource = 5.5$ mas yr$^{-1}$, and impact parameter $\imparam = 0.4$ Einstein radii. The $x$-axis in each panel is $\log{\sepparam}$, where $\sepparam$ is the binary separation in units of the Einstein radius. Top left: $q$ is varied on the $y$-axis. Bottom left: $\imparam$ is varied. Top right: $M$ is varied. Note that, since the Einstein radius is proportional to $\sqrt{M}$, the physical BBH separation will increase for increasing M and constant $\log{\sepparam}$. Bottom right: $\vsource$ is varied. Dotted curves correspond to weak (1.0), moderate (2.5), and strong (5.0) evidence levels for Bayes factor values. Solid colour lines correspond to BBH merger time in $10^3$ Gyr. Points in the plane with $\ln{B} \geq 5$ correspond to BBH systems that could theoretically be detected with high certainty, given ideal observations with KMTNet. Points in the plane with $\ln{B} \leq -5$ correspond to BBH systems that would appear to be single lenses with high certainty. Red stars in each panel correspond to the parameters of our fiducial model at the binary separation
$\sepparam = 0.0250$ Einstein radii, marking the threshold of strong evidence in support of binarity ($\ln B$ = 5).}
\end{figure*}

\section{\label{results}Results}
The test in Bayesian statistics of support for one model over another is the Bayes factor, which is the ratio of the binary-lens evidence to the single-lens evidence: $B \equiv E_B/E_S$. We apply the scale from \cite{jeffreys} to determine support, where $|\ln{B}| \geq 5$ is strong evidence, $2.5 \leq |\ln{B}| < 5$ is moderate evidence, $1 \leq |\ln{B}| < 2.5$ is weak evidence, and $|\ln{B}| < 1$ is inconclusive. In the limits that $q \rightarrow 0$ and $\sepparam \rightarrow 0$, the peak likelihood ratio $\rightarrow 1$ while $\ln{B}$ becomes negative, indicating that the single-lens model is favored.

We compute whether a BBH system with mass parameters similar to those of the first detected LIGO event, GW150914, would be detectable by microlensing in our own Galaxy. Our fiducial model is $M = 65M_{\odot},~q = 0.8$, $\vsource = 5.5$ mas yr$^{-1}$, and $\imparam = 0.4$ Einstein radii. We choose a lens distance $D_L = 4$ kpc and a source star in the Galactic Bulge at $D_S = 8$ kpc with apparent magnitude $I = 15$, \change{on the bright end of typical source magnitudes in current microlensing surveys.} For this model, we find that BBH separations down to approximately $\sepparam = 0.0250$ Einstein radii are detectable with strong evidence, given ideal observing conditions for one year around the peak of the event. \change{For a dimmer source star with $I = 17$, separations down to only $\sepparam = 0.0270$ Einstein radii are detectable.}

We also calculate the $\Delta \chi^2$ between the binary and single lenses, both for our fiducial separation $\sepparam = 0.01$ Einstein radii and the minimum detectable separation $\sepparam = 0.0250$ Einstein radii. For the fiducial case, $\ln B = -7.65$ corresponds to $\Delta \chi^2 = 1.36$; for the minimum detectable $\sepparam = 0.0250$ Einstein radii, $\ln B \simeq 5$ and $\Delta\chi^2 = 53.4$.

The approximate merger time of binary black holes is (see \cite{peters}):
\begin{equation}
\label{merge}
t_0 = \dfrac{5\sepparam^4c^5}{256\eta M^3G^3}
\end{equation}
where $c$ is the speed of light and $G$ is the gravitational constant. Therefore, BBHs that fit our fiducial model with a separation of $\sepparam = 0.0250$ Einstein radii will merge in about $2.08\times10^3$ Gyr, which is much greater than the 13.8 Gyr age of the Universe. BBHs formed with these parameters in the early Universe can only serve as present-day GW sources if their initial separation was $\sepparam \leq 0.00716$ Einstein radii, $\sim 3.5$ times smaller than the minimum separation detectable by KMTNet. The dependence of this minimum separation on the binary parameters is explored in Fig.~2, which displays Bayes factor contours for one year of ideal observations for our fiducial model in the $q$-$\sepparam,~\imparam$-$\sepparam,~M$-$\sepparam$, and $\vsource$-$\sepparam$ planes along with curves of constant merger time.

\section{\label{disc}Discussion}
The Bayes factor values we calculate are for ideal observing conditions over the course of one year, centred on the event peak. We do not explore the effects of finite-source size or other potential issues such as interstellar extinction or the difficulties of observing through Earth's atmosphere.

Our results indicate that KMTNet cannot detect binary modulation in the microlensing light curves of BBHs with separations small enough to merge through GW emission within the age of the Universe. The BBHs that can be identified have wider separations than these GW sources, potentially corresponding to systems that failed to experience a decrease in orbital separation during common-envelope evolution between the core collapses of the BBH stellar progenitors (see \cite{webbink}, \cite{dominikb}, \cite{gerosa}). Microlensing BBH searches thus complement GW based searches through their sensitivity to more widely separated BBHs.

The minimum separation for which the binary-lens model is favored can be reduced by decreasing the source distance $D_S$, decreasing the ratio $D_L/D_S$, decreasing the impact parameter $\imparam$, decreasing the proper motion $\vsource$, or increasing $q$. Larger total masses $M$ can also slightly improve detectability. To reduce this minimum separation from 0.0250 to 0.00716 Einstein radii (small enough to merge within the age of the Universe), a future experiment must increase the $\Delta\chi^2$ beyond that obtainable with KMTNet by a factor of $(0.025/0.00716)^4 \simeq 150$ since the binary signal $d_k - m_k$ is proportional to the binary quadrupole moment ($\propto {\sepparam}^2$) and $\chi^2 \propto (d_k - m_k)^2$. This could be achieved for example with continuous observations (10 minute cadence/120 s exposure = 5) over 10 years of the event by a network with 3 times as many telescopes as KMTNet ($5 \times 10 \times 3 = 150$).

As discussed in \cite{pennya} and \cite{pennyb}, there are some degeneracies of note with binary microlensing signals. For example, a binary stellar source with a single lens could mimic the astrometric signal of a binary lens, though the photometric signal would be discernibly different. We expect the centroid shifts due to binary sources to be much smaller than those due to binary lenses -- if the source's apparent magnitude is bright enough, \textit{Gaia}'s 24 microarcsecond precision (see \cite{esa}) could resolve binary star separations $< 1$ AU at the Galactic Centre. Another potential degeneracy could arise due to source stars with inherent pulsations when there is no lens present at all. Such pulsations would continue outside the timeline of a standard lensing event. Therefore, observations outside the expected lensing event time could help eliminate this degeneracy.

Calculating the rates of BBH microlensing events would require astrophysical modeling of BBH formation beyond the scope of this paper (see \cite{belczynski}). \cite{stefanob} estimated that current microlensing surveys can detect 0.38 black hole lenses per decade per deg$^2$. The fraction of these black holes in binary systems is highly uncertain but could be of order unity. KMTNet has a 4 deg$^2$ field of view and can look at 4 fields with a 10 minute cadence (\cite{henderson14}), yielding $\sim6$ black hole microlensing events per decade. \change{In addition, future space-based surveys like WFIRST will likely achieve increased photometric capability over KMTNet while still preserving the 10-minute cadence and $\sim1$ year survey duration, likely leading to more detections of black hole and BBH microlensing events.}

KMTNet is capable of identifying BBHs with high-cadence sampling of a microlensing event. However, as demonstrated in this paper, it does not possess the photometric precision necessary to observe microlensing events by BBHs with merger times within the age of the universe. We encourage analyses of current and future microlensing survey data to search for similar modulation in all long-duration events, providing a new channel for the discovery of short-period BBHs in our Galaxy.

\section*{Acknowledgements}
The authors acknowledge financial support from the Texas Space Grant Consortium Columbia Crew Memorial undergraduate scholarship, as well as from the University of Texas at Dallas Undergraduate Research Award. They would also like to thank \change{Matthew Penny and an anonymous referee} for their helpful suggestions regarding this manuscript. MK is supported by Alfred P Sloan Foundation Grant No. FG-2015-65299 and NSF Grant No. PHY-1607031. LJK is supported by NSF Grant No. AST-1517954, NASA Grant No. NNX16AF53G, and NASA HST-GO-12871.001-A.


\begin{thebibliography}{99}
\bibitem[\protect\citeauthoryear{Abbott et al.}{2016a}]{ligo}
Abbott B. P. et al., 2016a, \prl, 116, 061102

\bibitem[\protect\citeauthoryear{Abbott et al.}{2016b}]{ligo2}
Abbott B. P. et al., 2016b, \prl, 116, 241103

\bibitem[\protect\citeauthoryear{Abbott et al.}{2016c}]{ligo3}
Abbott B. P. et al., 2016c, \apjl, 818, L22

\bibitem[\protect\citeauthoryear{Belczynski et al.}{2016}]{belczynski}
Belczynski K., Repetto S., Holz D. E., O'Shaughnessy R., Bulik T., Berti E., Fryer C., Dominik M., 2016, \apj, 819, 108

\bibitem[\protect\citeauthoryear{Di Stefano \& Scalzo}{1999}]{stefano}
Di Stefano R., Scalzo R.~A., 1999, \apj, 512, 579

\bibitem[\protect\citeauthoryear{Di Stefano}{2008}]{stefanob}
Di Stefano R., 2008, \apj, 684, 59

\bibitem[\protect\citeauthoryear{Dominik}{1999}]{dominik}
Dominik M., 1999, \aap, 349, 108

\bibitem[\protect\citeauthoryear{Dominik et al.}{2012}]{dominikb}
Dominik M., Belczynski K., Fryer C., Holz D. E., Berti E., Bulik T., Mandel I., O'Shaughnessy R., 2012, \apj, 759, 52

\bibitem[\protect\citeauthoryear{Dubath, Gasparini, \& Durrer}{2007}]{dubath}
Dubath F., Gasparini M. A., Durrer R., 2007, Phys. Rev. D, 75, 024015

\bibitem[\protect\citeauthoryear{European Space Agency}{2014}]{esa}
European Space Agency, 2014, Science Performance. \url{http://www.cosmos.esa.int/web/gaia/science-performance} (accessed June 7, 2016)

\bibitem[\protect\citeauthoryear{Gerosa et al.}{2013}]{gerosa}
Gerosa D., Kesden M., Berti E., O'Shaughnessy R., Sperhake U., 2013, Phys. Rev. D, 87, 104028

\bibitem[\protect\citeauthoryear{Guo et al.}{2015}]{guo}
Guo X., Esin A., Di Stefano R., Taylor J., 2015, \apj, 809, 182

\bibitem[\protect\citeauthoryear{Han, Chun, \& Chang}{1999}]{han}
Han C., Chun M.-S., Chang K., 1999, \apj, 526, 405

\bibitem[\protect\citeauthoryear{Henderson et al.}{2014}]{henderson14}
Henderson C. B., Gaudi B. S., Han C., Skowron J., Penny M. T., Nataf D., Gould A. P., 2014, \apj, 794, 52

\bibitem[\protect\citeauthoryear{Henderson}{2015}]{henderson15}
Henderson C. B., 2015, \apj, 800, 58

\bibitem[\protect\citeauthoryear{Hobson et al.}{2010}]{hobson}
Hobson M. P., Jaffe A. H., Liddle A. R., Mukherjee P., Parkinson D., eds, 2010, Bayesian Methods in Cosmology. Cambridge Univ. Press, Cambridge, UK

\bibitem[\protect\citeauthoryear{Jeffreys}{1961}]{jeffreys}
Jeffreys H., 1961, Theory of Probability, 3rd edn. Oxford Univ. Press, Oxford, UK


\bibitem[\protect\citeauthoryear{Nucita et al.}{2014}]{nucita}
Nucita A.A, Giordano M., De Paolis F., Ingrosso G., 2014, \mnras, 438, 2466

\bibitem[\protect\citeauthoryear{O'Hagan}{1994}]{stats}
O'Hagan A., 1994, Kendall's Advanced Theory of Statistics, Vol. 2B. Wiley, Hoboken, NJ

\bibitem[\protect\citeauthoryear{Penny, Mao, \& Kerins}{2011a}]{pennya}
Penny M. T., Mao S., Kerins E., 2011a, \mnras, 412, 607

\bibitem[\protect\citeauthoryear{Penny, Kerins, \& Mao}{2011b}]{pennyb}
Penny M. T., Kerins E., Mao S., 2011b, \mnras, 417, 2216

\bibitem[\protect\citeauthoryear{Peters \& Mathews}{1963}]{peters}
Peters P. C. \& Mathews J., 1963, Phys. Rev., 131, 435

\bibitem[\protect\citeauthoryear{Sajadian}{2014}]{sajadian2014}
Sajadian S., 2014, preprint (arXiv:1401.6416)

\bibitem[\protect\citeauthoryear{Sajadian}{2015}]{sajadian2015}
Sajadian S., 2015, AJ, 149, 147

\bibitem[\protect\citeauthoryear{Schneider \& Weiss}{1986}]{schneider}
Schneider P., Weiss~A., 1986, A\&A, 164, 237

\bibitem[\protect\citeauthoryear{Trotta}{2008}]{trotta}
Trotta R., 2008, Contemp. Phys., 49, 71

\bibitem[\protect\citeauthoryear{Webbink}{1984}]{webbink}
Webbink R. F., 1984, \apj, 277, 355

\bibitem[\protect\citeauthoryear{Zheng \& Gould}{2000}]{zheng}
Zheng Z., Gould A., 2000, \apj, 541, 728
\end{thebibliography}
\end{document}